# Motor Learning Without Moving: Hand Localization after Passive Training

**Mostafa, AA[1,2]** | **'t Hart, BM[3]** | **Henriques, DYP[1,3]**

[1]Kinesiology and Health Science, York University, Toronto, Ontario, Canada

[2]Faculty of Physical Education, Mansoura University, Mansoura, Egypt

[3]Centre for Vision Research, York University, Toronto, Ontario, Canada

**Correspondence**
BM 't Hart, Bethune 357, 4700 Keele Street, Toronto ON, M3J 1P3 Canada
Email: thartbm@gmail.com

**Funding information**
NSERC Discovery Grant (DYPH), German Research Foundation (DFG) grant: HA 6861/2-1 (BMtH)

An accurate estimate of limb position is necessary for movement. Where we localize our unseen hand after a reach depends on felt hand position, or proprioception, but often only predicted sensory consequences based on efference copies of motor commands are considered. Both signals should contribute, so here we use passive training with rotated visual feedback of hand position to prevent updates of predicted sensory consequences, but still recalibrate proprioception. After this training we measure participants' hand location estimates based on both efference-based predictions and afferent proprioceptive signals with self-generated hand movements as well as based on proprioception only with robot-generated movements. The changes in hand localization are equally large after training with robot- and self-generated hand movements. Both motor and proprioceptive changes are only slightly smaller as those after training with self-generated movements, confirming that recalibrated proprioception contributes to motor learning. (data, preprint)

# 1 | INTRODUCTION

Sensory information is central to how we control all our movements. Our brain is even thought to use predicted sensory consequences derived from efferent copies of motor commands for motor control [1]. When training with rotated visual feedback of the hand, we update these predictions [2]. Additionally, such training leads to a recalibration of our sense of felt hand position - "proprioception" - to be more aligned with the distorted visual feedback [3]. Both of these changes have been measured by asking people to localize where their unseen hand is – before and after training [4, 5, 6, 7, 8]. While our lab has previously found that proprioception accounts for a large portion of the change in hand localization [9], it is far from clear how much each process contributes or how to tease them apart. Here we use passive training, that removes the need to update predicted sensory consequences, in an attempt to isolate the contribution of proprioception to hand localization.

The predicted sensory consequences of movements may play several important roles in motor learning and control. Predicted sensory consequences allow us to correct our movements before sensory error signals are available, they can be used to select movements that best achieve our goals and they may inform us on the location of our limbs. Hence measuring predicted sensory consequences is valuable in movement research. In visuomotor rotation adaptation tasks, the actual sensory outcome is systematically different from the expected outcome, so that participants update their predictions on the outcome of the trained movements. In previous experiments, people were asked to make a movement and then indicate the location of, or "localize," their unseen, right hand, before and after training with rotated visual feedback [5, 7]. Since there was no visual information available to the participants, the predicted sensory consequences of the movement should be used in localizing the unseen hand. Both studies found a significant shift in hand localization, providing evidence that predicted sensory consequences are indeed updated as a result of visuomotor rotation adaptation.

However, our lab has shown that our sense of where we feel our hand to be, proprioception, is also reliably recalibrated after visuomotor rotation adaptation [3, 9, 10, 11, 12, 13, 14, 15, 16, 17, 18]. This has also been shown by other labs [19] and a comparable proprioceptive change is induced with force-field adaptation [20]. As proprioception also informs us on the location of our limbs, we have on occasion used a task that is very similar to hand localization to investigate this [3, 6, 21, 22]. Although proprioceptive recalibration has been largely ignored as an explanation for changes in hand localization, we and others have shown that it accounts for a substantial part of the changes in localization, along with updates in predicted sensory consequences [9, 19]. Nevertheless, it is far from clear how much proprioception and prediction each contribute to hand localization.

Here we intend to further examine the contribution of proprioception to hand localization. To do this, we use passive training, where a robot arm moves the participant's arm out, so that the cursor always directly goes to the target [19, 23, 24]. This means there is no efference copy available and no visuomotor error-signal, both of which are required to update predicted sensory consequences. However, we impose a discrepancy between vision and proprioception that drives proprioceptive recalibration. Thus, changes in hand localization after this type of training should be due to proprioceptive recalibration only. We use the same experimental protocol as before [9], so that we can compare localization shifts between the two different types of training, and can better assess the contributions of predicted sensory consequences and proprioception to hand localization.

## 2 | METHODS

We set out to test the relative contributions of proprioception and efference-based prediction to hand localization. We use visual training with robot-generated hand movements to prevent updates of predicted sensory consequences, but still elicit proprioceptive recalibration.

### 2.1 | Participants

Twenty-five right-handed participants were recruited for this study. One participant was excluded for not following task instructions, and three were excluded for low performance on a task that ensures attention during the passive training. All analyses presented here pertain to the remaining twenty-one participants (13 females and 8 males, mean age: 20.1 +/- 2.3 years), but the data of the three low-performing participants is included in the online dataset [25]. All had normal or corrected-to-normal vision, and provided prior, written informed consent in accordance with the ethical guidelines set by the York Human subjects Review Subcommittee and received credit toward an undergraduate psychology course. Participants were screened verbally, and all reported being right handed and not having any history of visual, neurological, and/or motor dysfunction.

### 2.2 | Setup

Participants sat in a height-adjustable chair to ensure that they could easily see and reach all targets presented on a reflective screen (see Fig 1). During all tasks, they held the vertical handle on a two-joint robot manipulandum (Interactive Motion Technologies Inc., Cambridge, MA) with their right hand so that their thumb rested on top of the handle. A monitor (Samsung 510 N, refresh rate 60 Hz) was mounted 11 cm above the reflective screen, such that images displayed on the monitor appeared to lie in the horizontal plane where the right hand was moving. The reflective screen was mounted horizontally 18 cm above the robot manipulandum. A touch screen was mounted 13 cm underneath the reflective surface, so that subjects could indicate the location of the unseen right-hand locations (specifically the unseen thumb) with their visible left hand, which was lit up with a small spot light (only in localization tasks). The room lights were dimmed and the participants' view of their right hand was blocked by the reflective screen, as well as a dark cloth draped between the touch screen and participants' right shoulders.

### 2.3 | Procedure

The first part of the experiment used training with a cursor aligned with the hand and the second part had training with a cursor rotated around the start position (Fig 1b; white rows in Table 1). During the training with rotated feedback, the cursor was gradually rotated 30° clockwise. This introduced a discrepancy between the actual, or felt, hand position and the visual feedback, that should evoke proprioceptive recalibration. However, the movements were robot generated, so that there were no predicted sensory consequences based on the outgoing motor command. Hence the prediction errors that are thought to lead to motor learning were absent as well. After both types of training, participants did open-loop reaches as well as two kinds of hand localization tasks, to test the effect of training on proprioceptive and predictive hand estimates.

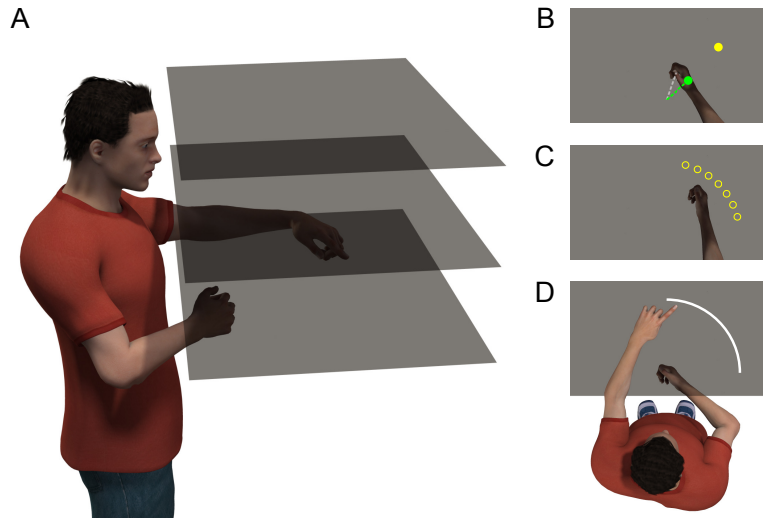

**FIGURE 1** Setup and tasks: a) Participants moved their unseen right hand with visual feedback on hand position provided through a mirror (middle surface) half-way between their hand and the monitor (top surface). A touchscreen located just above the hand was used to collect responses for the localization tasks (bottom surface). b) Training task. The target, shown as a yellow disc, is located 10 cm away from the home position at 45°. In the rotated training tasks, the cursor (shown here as a green circle) represents the hand position rotated 30° relative to the home position. c) No-cursor reach task. Targets are located 10 cm away from the home position at 15°, 25°, 35°, 45°, 55°, 65°, and 75°, shown by the yellow circles here (only one was shown on each trial). While reaching to one of these targets, no visual feedback on hand position is provided. d) Localization task. The participants' unseen, right hands moved out, and subsequently participants indicated the direction of the hand movement by indicating a location on an arc using a touch screen with their visible left index finger.

### 2.3.1 | Exposure training

In what we call 'exposure training' the participants did not move their hand toward the target, but the robot did. In this task (Table 1), the right hand was represented by a cursor (green disk, 1 cm in diameter, Fig 1b) located directly above participant's thumb. The robot moved the participant's unseen right hand (and the cursor) along a direct path toward a visual yellow target disk and back to the starting position (1 cm in diameter, Fig 1b). The home position was located approximately 20 cm in front of participants and the visual target located 10 cm from the home position at 45° (Fig 1b). In order to make sure participants were paying attention to the cursor, the cursor was switched off for 2 screen refreshes (~33.3ms) on 50% of the trials at a random distance between 4 and 9 cm from the home position and participants were asked to report this using a button press with the left hand. Performance on this task was used to screen participants.

During the first half of the experiment, the cursor and hand path were aligned during exposure training. In the second part of the experiment, the "rotated" session, the same visual training target at 45° was used, and the cursor kept moving straight to this target. However, the robot-generated hand path gradually rotated 30° CCW (Fig 1b) with respect to the visible target and the cursor in increments of 0.75°/trial, so that the full rotation was reached after 45 trials. This mimicked error-free responses to a gradual visuomotor rotation of 30° CW. The initial training consisted of 50 trials in the aligned part and 90 in the rotated part. In between open-loop reach tasks and localization tasks (Table 1) extra training tasks were done, each of which consisted of 10 trials in the aligned part of the experiment and 60 trials in the rotated part (Table 1).

**TABLE 1** Task order.

| # trials | | training type | |
|---|---|---|---|
| aligned | rotated | **Exposure training** | **Classic training** |
| 50 | 90 | training | training |
| - | 21 | no-cursor | no-cursor |
| - | 60 | training | training |
| 25 | 25 | active delayed localization | active delayed localization |
| 21 | 21 | no-cursor | no-cursor |
| 10 | 60 | training | training |
| 25 | 25 | passive delayed localization | passive delayed localization |
| 21 | 21 | no-cursor | no-cursor |
| 10 | 60 | training | training |
| 25 | 25 | active delayed localization | *(active online localization)* |
| 21 | 21 | no-cursor | no-cursor |
| 10 | 60 | training | training |
| 25 | 25 | passive delayed localization | *(passive online localization)* |
| 21 | 21 | no-cursor | no-cursor |

### 2.3.2 | No-cursor reaching

The trials in no-cursor reaching (Table 1, light gray rows) serve as a classical measure of motor adaptation. On each of these trials participants were asked to reach with their unseen right hand to one of 7 visual targets, without any visual feedback of hand position. The targets were 10 cm from the home position, located radially at: 15°, 25°, 35°, 45°, 55°, 65°, and 75° (Fig 1c). A trial started with the robot handle at the home position and, after 500 ms, the home position disappeared and the target appeared, cuing the participants to reach for the target. Once the participants thought they had reached the target they held their position for 250 ms, and the target and the home position disappeared, cuing participants to move back to the home position along a straight, constrained path, to begin the next trial. If participants tried to move outside of the path, a resistance force, with a stiffness of 2 N/(mm/s) and a viscous damping of 5 N/(mm/s), was generated perpendicular to the path. In every iteration of the no-cursor reach task, each target was reached three times, for a total of 21 reaches in pseudo-random order. The no-cursor reaching task was performed four times in the aligned part of the experiment and five times in the rotated part of the experiment.

### 2.3.3 | Localization

In this task (Fig 1d; Table 1, dark gray rows) participants indicated where they thought their unseen right hand was after a movement. First, an arc appeared, spanning from 0° to 90° and located 10 cm away from the home position and the participants' unseen, right hand moved out from the home position in a direction towards towards a point on the arc. The hand was stopped by the robot at 10 cm from the home position and then, to prevent the online proprioceptive signals from overriding the predictive signals [5, 9], the hand was moved back to the home position using the same kind of constrained path as used for the return movements in the no-cursor task. Participants indicated with the index finger of their visible, left hand on the touch screen mounted directly above the robot handle where they thought their trained hand had crossed the arc.

Crucially, there were two variations of this task. First, in the 'active' localization task participants generated the movement themselves, as they could freely move their unseen right hand from the home position to any point on the arc. Second, there was a 'passive' localization task where the robot moved the participants' hand out and back, to the same locations the participants moved to in the preceding 'active' localization task in a shuffled order (hence, active localization is done first). In active localization, participants have access to both proprioceptive information as well as an efference-based prediction of sensory consequences, but in passive localization, only proprioception should be available. The active and passive localization task each consisted of 25 trials, and each of the tasks was done a total of four times; twice after aligned and twice after rotated training.

### 2.3.4 | Classic training

The paradigm described above is an exact replica of a paradigm we used earlier [9] with two exceptions. First, we used exposure training here, instead of the standard reach training with volitional movements, which we will call 'classic' training. Second, all localization is delayed until the right hand has returned to the home position in this study (see Table 1), so that instead of both delayed and online localization we have two repetitions of each delayed localization task. With this paradigm we can compare changes in localization and no-cursor reaches change after exposure training with changes in the same measures after classic visuomotor adaptation training.

### 2.4 | Analyses

Prior to any analyses, both the localization responses and the no-cursor reach data were visually inspected and trials where the participants did not follow task instruction were removed (e.g. several movements back and forth, or a touch-screen response on the home position, instead of on the white arc).

### 2.4.1 | Localization

Localization responses were taken as the (signed) angular difference between vectors through the home position and the actual hand position as well as the location indicated on the touch screen. Prior to analyses, idiosyncratic differences in performing this task were countered. Before conversion to degrees angle, a circle with a 10 cm radius was fit to the touch screen responses of each participant and the offset of this circle's centre was subtracted from all response coordinates, so that all responses fell close to the arc. Then, a smoothed spline was fit to every participant's response errors in each of the four localization tasks (aligned vs. rotated and active vs. passive) and these were used to obtain localization errors at the same locations used for the no-cursor reaches (15°, 25°, 35°, 45°, 55°, 65° and 75°), but only if that location fell within the range of the data (i.e. we only interpolate). This way localization responses could be compared across participants despite the freely chosen reach directions. At the 15° location 7/21 participants didn't have an estimate in one or more of the four localization tasks (in the "classic" data it was also 7/21). While that data is shown in the figures, we did not use it for analysis.

First we test if localization responses shifted following rotated exposure training compared to aligned. We then test if the shift in localization responses is different for active and passive localization, and we run analyses comparing localization after exposure training with localization after "classic" training. Finally, we explore the generalization of localization responses and if they are different between the groups doing classic and exposure training.

### 2.4.2 | Reach aftereffects

To assess any reach adaptation that may have occurred after exposure training we analyzed reach endpoint errors in no-cursor trials. Reach endpoint errors were the (signed) angular difference between a vector from the home position to reach endpoint and a vector from the home position to the target. We obtained reach aftereffects by subtracting reach endpoint errors after aligned training from those after rotated training. No-cursor endpoint errors were analyzed to see if participants adapted the direction of their reaching movements after rotated exposure training. We also tested if any such change decayed, i.e. if it was the same immediately after exposure training, or when a localization task was done in between exposure training and no-cursor reaches. Furthermore, we tested if the generalization of reach aftereffects is different between exposure and classic training and if there is any generalization of reach aftereffects.

Pre-processing and analyses were done in R 3.4.4 [26], using lme4, lmerTest and various other packages. Most analyses used linear mixed effects models, since there is some missing localization data. These were "converted" to more readable ANOVA-like output, using a Satterthwaite approximation [27]. Highly similar results were obtained with a Chi-square approximation. Data, scripts and a notebook with analyses have been made available on the Open Science Framework [25] (https://osf.io/zfdth/).

In short, this experiment allowed us to test how mere exposure to a visual-proprioceptive discrepancy changes both reach aftereffects and hand localization responses, and compare them with those obtained after more regular, "classic" training.

## 3 | RESULTS

In this study we intend to further elucidate the relative contributions of (updated) predicted sensory consequences and (recalibrated) proprioception to hand localization. We can parcel out these contributions by measuring hand localization after both robot-generated and self-generated movements. Finally we compare the data from the current experiment with those obtained in an earlier study that used an identical paradigm, but trained with self-generated movements, or "classic" training.

### 3.1 | Localization

Here we test our hypothesis that exposure training does not lead to changes in predicted sensory consequences. Since the difference between active and passive localization stems only from the presence or absence of predicted sensory consequences, there should be no difference between the two if predicted sensory consequences are not changed by exposure training. At first glance, it seems there might be a difference between active and passive localization shifts in exposure training (see Fig 2a), although it is smaller than in classic training (Fig 2b).

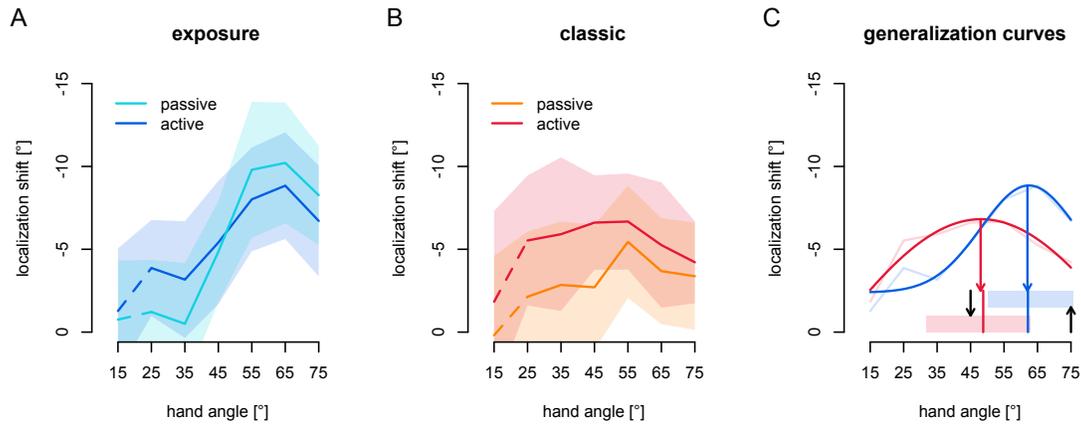

**FIGURE 2** **Hand localization:** The shifts of the angles of touchscreen responses in all variations of the localization task, using spline-interpolated estimates for hand angles matching the reach targets in the no-cursor reach block. **a)** Localization shifts after exposure training. Dark blue: active localization shifts, Light blue: passive localization shifts. **b)** Localization shifts after classic training. Dark red: active localization shifts, Orange: passive localization shifts. The dashed line segments illustrate that the 15° data is not used for statistical analyses (see Methods). **c)** Generalization curves of active localization shifts after exposure training (blue) and classic training (red). Shaded areas: 95% confidence intervals for the peak of the generalization curve (red and blue lines through shaded area indicate 50% points). Downward black arrow: visual trained target. Upward black arrow: hand location during training.

To test if rotated exposure training induces changes in hand localization, we fit an LME to the localization errors throughout the workspace using session (aligned or rotated), movement type (active and passive) and hand angle (25°, 35°, 45°, 55°, 65° and 75°), and all interactions as fixed effects and participant as random effect. There was an effect of session ($F(1,450.5)=155.8$; $p<0.001$), showing that exposure training leads to changes in hand localization. There was also an effect of hand angle ($F(5,450.6)=6.54$; $p<.001$) and an interaction between hand angle and session ($F(5,450.3)=8.25$; $p<.001$), which we'll explore below, but no other effects (all $p>.60$). Since localization responses did shift, we use the difference between hand localization after rotated training and after aligned training (as plotted in Fig 2) for further tests.

If this shift in localization after exposure training partly reflects predicted sensory consequences, then shifts in active localization, that rely on both (recalibrated) proprioception and (updated) predictions should be different from shifts in passive localization that only rely on (recalibrated) proprioception. We fit an LME to the change in localization using movement type (active or passive localization) and hand angle, as well as their interaction as fixed effects and participant as random effect. There was no effect of movement type ($F(1,211.8)=0.07$; $p=0.79$). There was an effect of hand angle ($F(5,212.2)=10.8$; $p<0.001$), but no interaction between hand angle and movement type ($F(5,211.8)=1.23$, $p=.29$). The lack of an effect of movement type suggests that predicted sensory consequences did not contribute to localization in this paradigm.

In order to compare hand localization shifts after exposure training with those after classic training [9], we fit an LME to localization shift using training type (exposure vs. classic), movement type (active vs. passive) and hand angle and all interactions as fixed effects and participant as random effect. There was a main effect of movement type ($F(1,422.0)=6.22$; $p=.013$) and of hand angle ($F(5,422.7)=8.19$; $p<.001$), as well as an interaction between training type and hand angle (F(5,422.7)=4.54; p<.001) and between training type and movement type (F(1,422.0)=4.48, p=.035), but there was no main effect of training type (F(1,39.1)=0.92, p=.34) and no other effects (all p>.14). These results also suggest that the magnitude of the shifts in localization are comparable between classic and exposure training, but that the pattern of generalization is different.

To address our main question, we will look at the interaction between training type (exposure vs. classic) and movement type (active vs. passive) we found above. Since there is no difference between active and passive localization shifts after exposure training alone, the interaction between training type and movement type should be caused by an effect of movement type on the localization shifts after classic training, as we found previously [9]. This means that shifts in hand localization after exposure training indeed rely on recalibrated proprioception alone, while after classic training, there also is a contribution of predicted sensory consequences to active localization.

For completeness, we explore the potentially different generalization patterns of localization shifts after classic and exposure training (Fig 2c). The LME indicates no difference in overall amplitude of localization shifts between the groups, so the interaction between training type and hand angle might stem from a generalization that does not peak at the trained location after exposure training. Using the active localization shifts only (which are larger, and arguably more similar to reach aftereffects), we bootstrap a 95% confidence interval for the peak localization shift across participants in each group. Here we include the data at 15° where it is available. After classic training, the peak localization shift was at 48.8° (95% confidence: 31.6° - 62.8°; red area in Fig 2c), and after exposure training the peak localization shift was at 62.1° (95% confidence: 50.0° - 75.6°; blue are in Fig 2c). This means that peak localization after classic training is lower than after exposure training, but not vice versa. Also note that the confidence interval for the peak localization shift after classic training includes the trained target (45°), but not after exposure training [5]. In short, the LME for localization shifts indicates a different generalization curve after exposure and classic training, which might be partially explained by a different position of the peak localization shift after exposure or classic training.

## 3.2 | Reach aftereffects

Apart from proprioception and prediction, we want to see if rotated exposure training has any effect on open-loop reaches and if these are robust. We measure whether participants adapted their reach directions by assessing their reach errors in no-cursor reach trials after aligned and rotated exposure training. In Fig 3, the changes in no-cursor endpoint errors, or reach aftereffects, appear to be well over 5°. First, to test if exposure training affects open-loop reach direction, we fit a linear mixed effects model (LME) to reach endpoint error using session (aligned; all blocks, or rotated; only the first block immediately after training) and target (15°, 25°, 35°, 45°, 55°, 65° and 75°), as well as their interactions as fixed effects and participant as random effect. There is an effect of session (F(1,260)=93.81, p<.001), that is: exposure training leads to substantial reach aftereffects. There was no effect of target (F(6,260)=1.07, p=.37) and no interaction (F(6,260)=1.00, p=.42). Since there is an effect of session, we now take the differences in reach endpoint errors between the rotated and aligned session for every participant and target as reach aftereffects, and use those for all further analyses.

To see if reach aftereffects decayed during the localization tasks, we compared reach aftereffects in the initial no-cursor block, that immediately followed training, with those in the later blocks that followed a localization task. We fit an LME to the reach aftereffects with iteration (initial vs. later no-cursor blocks) and target (as above) as well as their interaction as fixed effects and participant as random effect. There is no effect of iteration (F(1,260)=2.72, p=0.10). There was an effect of target (F(6,260)=6.29, p<.001) but no interaction (F(6,260)=0.58, p=0.74). Hence, reach aftereffects were not appreciably different right after training and after the localization tasks. In other words, there was likely no noticeable decay of reach aftereffects during the localization tasks, so that we can collapse the data across

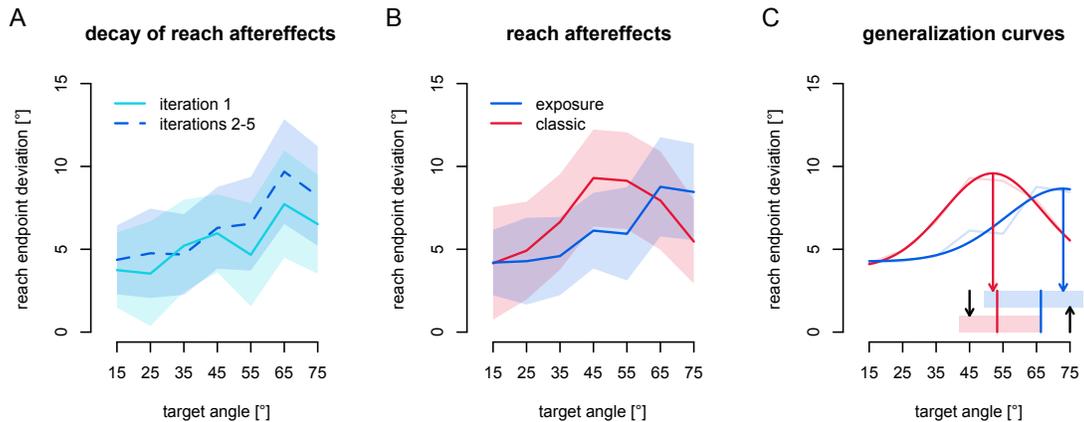

**FIGURE 3** **Reach aftereffects:** Changes of the angle of reach endpoints in the no-cursor tasks. **a)** Reach aftereffects across the experiment. Light blue: first no-cursor task in the rotated session (immediately following training), Dark blue, dashed line: the other four repetitions of the task (with localization in between training and no-cursor tasks). **b)** Reach aftereffects after classic and exposure training. Blue: exposure training, Red: classic training. **c)** Generalization curves of reach aftereffects after exposure training (blue) and classic training (red). Shaded areas: 95% confidence intervals for the peak of the generalization curve (red and blue lines through shaded area indicate 50% points). Downward black arrow: visual trained target. Upward black arrow: hand location during training.

iterations.

Next we compare the reach aftereffects after classic training with those after exposure training (Fig 3b). It appears as if there is little overall difference in magnitude, but there might be a shift of the generalization curve. We fit an LME to reach aftereffects with training type (classic vs. exposure), target (as above) as well as their interaction as fixed effects and participant as random effect. There is no main effect of training ($F(1,40)=0.11$, $p=.74$), indicating approximately equal magnitude of reach aftereffects after the two training types. There is an effect of target ($F(6,240)=8.36$, $p<.001$), indicating that reach aftereffects exhibit some form of a generalization curve. There is also an interaction between training type and target ($F(6,240)=2.27$, $p=.038$), indicating these generalization curves are different after the two training types.

We will explore these potentially different generalization curves here. In Fig 3c we can observe that reach aftereffects after exposure training seem not to peak at the trained target direction of 45° but a more forward direction. We test this by taking the 95% confidence interval of the centre of a normal curve fit to this data, bootstrapped across participants, and find that the median peak of the generalization curve of reach aftereffects after exposure training is at 66.3°, with a 95% confidence interval ranging from 49.3° to 78.9°. This would indeed suggest that the reach aftereffects after exposure training do not generalize around the trained direction of 45°, but at a more counter clockwise location. For classic training, generalization of reach aftereffects peaks at 53.2°, with a 95% confidence interval spanning 42.1° to 66.5°. So for classic training the 95% confidence interval for peak reach aftereffects does include the trained target. These confidence intervals also indicate that generalization of reach aftereffects does not peak at different target position after exposure and classic training. However, we can also observe that the full curve was not sampled after exposure training, so that curve fitting is not optimal. This means that – given our data – the interaction between target and training type found in the LME above can't be explained by a shifted generalization curve. This may be because our experiment was not set up to test this, and the similar angles where generalization peaks in both localization and reach aftereffects, suggests there may be a difference in where proprioception and prediction generalize strongest.

In summary, our main hypotheses are confirmed; exposure training leads to shifts in hand localization that are not different for active or passive localization, while movement type does have an effect on localization shift after classic training. Exposure training also causes robust reach aftereffects that are of comparable size to those found with classic training. There is some evidence that the generalization of both localization shifts and reach aftereffects are different after the two training types, and it appears this can partially be explained by a different peak of the generalization curves, but our data and analyses are not definitive.

## 4 | DISCUSSION

The position of limbs is important for planning and evaluating movements, and can be estimated through predicted sensory consequences, as well as visual and proprioceptive feedback. As in a previous study [9] here we quantify the contributions of predicted sensory consequences and proprioceptive recalibration to where we localize our hand after training with altered visual feedback of the hand. In classical adaptation paradigms, both predictions are updated and proprioception is recalibrated. Predictions are updated when they don't match actual sensory consequences, and proprioception is recalibrated when it doesn't match visual feedback. In this study we use "exposure" training, where the participants do not have volitional control of their movements. By design, this should eliminate efference copies and prevent updating predicted consequences of movements, but since the proprioceptive and visual feedback is the same, exposure training still allows proprioceptive recalibration. Before and after training, participants localize their hand, both after "active," self-generated movements that allow using predicted sensory consequences, and after "passive," robot-generated movements that only allow using proprioception. We calculate the training-induced shift in both types of localization given the same actual hand position. After classical training we previously reported larger shifts in active localization as compared to passive [9]. As we expected, after exposure training there are substantial shifts in localization, but no difference between active and passive localization, indicating that predictions are not updated after exposure training. Furthermore, we find that exposure training evokes substantial and robust reach aftereffects, indicating that recalibrated proprioception is used to plan movements.

Our lab previously investigated proprioceptive recalibration and reach aftereffects following visuomotor adaptation with classic training and matched exposure training. There we also found that proprioceptive recalibration is of similar magnitude in both training paradigms, but unlike here, reach aftereffects are usually much larger with classic training [13, 18, 22, 23, 24]. And while proprioceptive recalibration and reach aftereffects do proportionally increase with gradual increases in rotation size for classical training, they do not for exposure training [24]. The similar magnitude of proprioceptive recalibration and reach aftereffects following exposure training, but not classical training, suggest that this sensory recalibration is partly driving this modest change in movements. The effect of exposure training on movements is also demonstrated by savings and interference from exposure training to subsequent classic training [28] and transfer of exposure training effects from one hand to the other [29]. In the current study, we further demonstrate that exposure training affects movements and proprioception, but also measure its potential effect on predictive estimates.

Results similar to what we find here were reported in a study by Cameron and colleagues [19], using gain modulation of visual feedback of single-joint hand movements around the elbow. Their within-subjects experiment included both training with volitional movements as well as with passive movements and also tested perception of movements that were either passive or active. They too found a robust change in passive perception of hand movement (using a different measure), and these changes did not differ between the two types of training. Similarly, they found shifts in what we might call "active localization," although the task is different, after both training types. Like here, these shifts are larger after classic training as compared to exposure training. They also found that passive exposure leads to reach aftereffects, although these were smaller than those produced following "classical" training with altered visual gain. Both our findings, and those of Cameron et al. [19] indicate that updating predicted sensory consequences requires volitionally controlled movements that lead to prediction errors, while proprioception recalibrates equally in both types of training, and that recalibrated proprioception affects open-loop reaches. Our combined results suggest that updates in predicted sensory consequences only provide a partial explanation for motor learning.

Two related concerns about exposure training and passive localization are that the movements are not fully passive, so that efference-based predictions are still generated or that predicted sensory consequences are generated through another route. Cameron et al. [19] measured muscle activity (EMG) during passive movements and found no difference with stationary baseline muscle activity. This suggests that any movements generated in a passive condition are subthreshold, minimizing efference-based predictions. The brain areas generating predicted sensory consequences could also rely on afferent signals. However, such afferent signals are present in both active and passive movements, and if they would result in the same predictions, there would be no difference between active and passive localization after classic training, and no difference between the effects of exposure and classic training, and we find both are different. Hence, while we can not fully exclude any predictive signals in passive localization or exposure training, our data shows that any residual predictive signals in the passive movements we used are qualitatively very different from normal efference-based predicted sensory consequences.

In our classical training group, we not only see shifts for passive localization but even larger shifts for active localization which is consistent with a change in both proprioception and an update in predictions. In our exposure training, we did not find a consistent difference between active and passive localization, and none at the trained direction. Assuming that predictions were not updated in exposure training, a maximum likelihood estimate (MLE) or "optimal integration" [30] would predict that active localization should shift less than passive localization after exposure training. But of course this is not the case in our findings (although it is the case for Cameron et al.). This suggests that perhaps these two signals are not optimally integrated which is consistent with our comparisons of the variance between passive and active localization. In 't Hart and Henriques [9], we tested the prediction derived from MLE that hand localization with two signals – proprioception and prediction in active localization – should be more reliable, i.e. have lower variance, than hand localization with only one signal – proprioception only in passive localization. However, we found no difference in variance between active and passive localization, and recently replicated this in a much larger dataset [31]. Taken together, this suggests these different sources of information about unseen hand location are not optimally integrated. While localizing the unseen hand is less precise than locating (pointing to) a remembered visual target or a seen and felt hand location, we find that these bimodal estimates are rarely integrated optimally [32, 33, 34], although others have [35]. A more recent study [36] has also shed doubt on whether "optimal" or "Bayesian" integration is used for locating the hand with two afferent signals. Analogously, here we again can't find evidence that afferent and efferent information combine as a maximum likelihood estimate.

It seems clear that the cerebellum plays a role in motor learning as it appears to compute predicted sensory consequences, i.e. it implements a forward model [37, 38, 39]. People with cerebellar damage do worse on motor learning tasks [40, 41, 42, 43], and show decreased shifts in hand localization tasks following motor learning [5, 7]. This highlights that the cerebellum, and likely predicted sensory consequences, are important for motor learning, but does not explain the remaining shifts in hand localization. We previously found that proprioceptive recalibration is intact in people with mild cerebellar ataxia and that it is similar following exposure and classical training with a gradually introduced cursor rotation [18]. The remaining changes in hand localization found in cerebellar patients can be attributed to recalibrated proprioception which should be intact [18]. Analogously, here we show that in a paradigm that stops updates of predictions of sensory consequences, as supposedly in people with cerebellar damage, we still see substantial shifts in localization. Again, the remaining localization shifts can be explained if, along with predictions, the human brain uses afferent signals: recalibrated proprioceptive estimates, to localize the hand.

## 4.1 | Generalization

We do find some evidence that, after exposure training, the generalization curves for localization shifts are not centred on the visual location of the trained target; they don't peak at 45° but at 62°. In contrast, after classic training the peak of the generalization curve does peak close to the training target. It is possible that proprioceptive recalibration is not anchored to the visual goal of the training task as it is not a requirement to feel your hand at any specific point; rather, in classic training the visual cursor has to be brought to a visual target to end a trial. While in exposure training the movements are executed for the participants without error, their task is still to pay attention to the visual cursor while it moves to the visual target; there are no task demands on proprioception. Although we can't substantiate this here, the generalization curves of the reach aftereffects seem to mimic the generalization curves of localization shifts, suggesting a relationship between changes in state estimates and changes in movements. This needs to be tested further, but if these effects are true, they may provide insight into how state estimates are used to produce movements, and also may lead to a new method to disentangle the influence of recalibrated proprioception and more traditional updated internal models on motor changes, such as reach aftereffects. Either way, even though the experiment was not designed to investigate this, the shifted generalization curves of changes in localization after classic and exposure training suggest they are generated by different mechanisms. This is in line with our earlier findings that proprioception generalizes differently from reach adaptation [11].

## 4.2 | Conclusion

To sum up, after a training paradigm designed to prevent updating of predicted sensory consequences but allow recalibration of proprioception, we find substantial changes in where people localize their hand. This means that recalibrated proprioceptive estimates can explain shifts in hand localization. The distinct change in the direction of open-loop reaches we observe here, suggests recalibrated proprioception can contribute to motor adaptation in other contexts as well. Finally, we have some evidence that after exposure training, the shift in hand localization does not generalize around the trained target location. All of this confirms that different mechanisms underlie proprioceptive recalibration and motor adaptation.

**ACKNOWLEDGEMENTS**

We thank Shanaathanan Modchalingam for assistance in collecting the data.